\documentclass[aps, pre, reprint]{revtex4-1}

\usepackage{latexsym}
\usepackage{hyperref}
\usepackage{graphics}
\usepackage{graphicx}
\usepackage{url}
\usepackage{listings}
\usepackage{enumerate}
\usepackage{longtable}
\usepackage{textcase}
\usepackage{natbib}
\usepackage{color}
\usepackage{amsfonts}
\usepackage{amssymb}
\usepackage{amsmath}
\usepackage{amsthm}
\usepackage{boxedminipage}
\usepackage{bm}
\usepackage{booktabs}

\usepackage{lineno}
\usepackage{fullpage}

\newcommand{\br}{\vspace{0.3cm}}

\begin{document}
\title{ Phase Transition for the Chase-Escape Model on 2D Lattices}
\date{\today}
\author{Si \surname{Tang}}
\email[Email: ]{si.tang@duke.edu}
\affiliation{Department of Mathematics, Duke University, Durham, NC 27705, USA}
\author{George \surname{Kordzakhia}}
\email[Email: ]{george.kordzakhia@fda.hhs.gov}
\affiliation{U.S. Food and Drug Administration, Silver Spring, MD 20993, USA}
\author{Steven P. \surname{Lalley}}
\email[Email: ]{lalley@galton.uchicago.edu}
\affiliation{Department of Statistics, The University of Chicago, Chicago, IL 60637, USA}

\begin{abstract}
Chase-Escape is a simple stochastic model that describes a predator-prey
interaction. In this model, there are two types of
particles, red and blue. Red particles colonize adjacent empty sites at an exponential 
rate $\lambda_{R}$, whereas blue particles take over adjacent
red sites at exponential rate $\lambda_{B}$, but can never colonize
 empty sites directly. Numerical simulations suggest that there is a critical value $p_{c}$ for the
relative growth rate $p:=\lambda_{R}/\lambda_{B}$. When $p<p_{c}$, 
mutual survival of both types of particles has zero probability, and when $p>p_{c}$ mutual survival occurs with
positive probability. In particular, $p_{c} \approx 0.50$ for the square lattice case ($\mathbb Z^{2}$). Our simulations provide a plausible explanation for the critical value. 
Near the critical value, the set of occupied sites exhibits a fractal nature, and 
the hole sizes approximately follow a power-law distribution. 
\end{abstract}

\maketitle

\section{\label{sec:intro}Introduction}

The identification of mechanisms that allow for coexistence of
competing species is a fundamental problem of ecology. It is well
understood that mean-field models incorporating many different types
of interaction (predator-prey, competition, mutualism, etc.) commonly
have stable equilibria -- depending, of course, on the parameters
describing the strengths of the interactions -- with several species
coexisting.  However, such models assume both large populations and a
high degree of mixing, assumptions which are not always
appropriate. Stochastic models that attempt to account for population
sparsity and local interactions are more difficult to study
mathematically. In such models, space is often represented as a
$d-$dimensional lattice whose sites can be occupied by only a bounded
number of individuals (in the simplest models, only one per site), and
interactions are restricted to individuals at neighboring sites.  See
\cite{durrett:1999} for a review of stochastic interacting particle
system models in spatial ecology, and
\cite{Haggstrom:1998eh,Kordzakhia:2005ik, Kordzakhia:2005vu,
durrett-neuhauser,lanchier-neuhauser,durrett-schinazi} for studies of
some particular models that incorporate inter-species competition, and
\cite{Liggett:1985vl,Liggett:vl1999} for a broad overview of the
mathematical theory of stochastic interacting systems. Similar models
are commonly employed in the study of epidemic propagation, see for
instance, \cite{Aoki:2008iz, Alonso:2006ja, Ahn:2006kb, Argolo:2011jy,
deSouza:2011fr,lalley:2009, lalley-zheng:2010, Tome:2010jp,
Neri:2011ti, Sugiura:2009tt}. Critical phenomena for a number of
related models have been investigated in \cite{Antal:2001eo, Peltomaki:2005df, Hasegawa:2011hu, Herbut:2006ja, ZhiZhen:2009jv}.

There are few rigorous results concerning coexistence in two-type or
multi-type lattice models: \cite{durrett-neuhauser,
durrett-schinazi,Haggstrom:1998eh,Kordzakhia:2005ik} have shown that
for certain types of interaction and at certain parameter values
species coexistence is possible, but the mathematical techniques used
in these articles are highly model-specific.  In this paper, we report
on a simple two-type predator-prey model, called
``\textit{Chase-Escape}'', for which there seems to be little hope of
obtaining rigorous results. In this model, two types of particles,
predators (``\textit{blue}'') and prey (``\textit{red}''), interact on
a graph according to the following rules. At any time $t$, a vertex
can be occupied by at most one particle, blue or red. The system
evolves in continuous time as follows: (1) red particles colonize
 adjacent empty sites at exponential rate $\lambda_{R}$; (2) blue particles
take over  adjacent red sites at exponential rate $\lambda_{B}$, but can never
occupy empty sites directly; and (3) once a site is occupied by a
blue particle, it remains blue forever. Long-run properties of the
system are completely determined by the ratio
$p=\lambda_{R}/\lambda_{B}$ of the colonization rates.  The question of interest 
is when the two types of particles can co-survive forever. Since blue particles never go extinct,
this is equivalent to determining when the red particles can escape extinction.

Chase-Escape  has been studied on the 
complete graph case \cite{kortchemski2015predator}, on the $d$-ary homogeneous tree \cite{Kordzakhia:2005vu}
and on random Galton-Watson trees \cite{bordenave}.
In all these cases, the probability of survival of the red
particles is monotone in $p$, and there is a threshold value $p_{c}$
above which mutual survival can occur, but below which it cannot. 
It is natural to conjecture that this is also the case on lattices.  
However, there is no monotone coupling of the system at different parameter values, as
for other stochastic particle systems (e.g., the contact process) that
exhibit threshold behaviors, so the existence of such a threshold
on lattices remains unproved. It is not hard to see that whenever red particles grow faster than the blue particles, coexistence is possible \cite{coexistence-2018}; this implies that $p_{c}$, if exists, should be no greater than 1. In fact, on $\mathbb Z^{2}$, the critical value $p_{c}$ has been  conjectured by
Kordzakhia in 2003 and by James Martin to be strictly below 1, i.e., coexistence is possible even if 
red particles grow strictly slower than the blue particles. A recent paper \cite{coexistence-2018}
confirmed that coexistence is indeed possible for Chase-Escape on $\mathbb Z^{d}$ for some $d$ large and 
a different choice of red and blue colonization times where red particles spread slower than the blue particles.

Our simulations indicate that the probability of mutual survival is indeed monotone in $p$ 
not only for the square lattice $\mathbb Z^{2}$, but for the triangular, hexagonal, and 8-directional (Fig. \ref{fig:lattice}) lattices. Moreover, for the square lattice, the simulations strongly suggest that the
critical value is $p_{c}\approx 1/2$, confirming the conjecture that $p_{c} < 1$. Near the critical value, we find that the shape of occupied sites is fractal, which is extremely different from the case away from the critical value.

The rest of the paper is organized as follows: in
Sec. \ref{sec:methods}, we describe the Chase-Escape model on 2D
lattices in detail and compare it with the Richardson, Competition and
Escape models. We briefly describe our simulation method.  In Sec. \ref{sec:result}, we report the
critical values of $p$ when mutual survival becomes probable for the
four types of lattices. We then discuss the fractal properties of the
occupation zone near $p=p_{c}$, which confirms the criticality of $p_{c}$. When
presenting these results, we always report the square lattice case
first and in detail, and then briefly state the result on the other
three types of lattices.  In Sec. \ref{sec:conclusion}, we conclude
and discuss our results. 

\section{\label{sec:methods}Methods} \subsection{\label{subsec:model}
The Construction of The Chase-Escape model on 2D lattices} 
\begin{figure}[ht]
\begin{minipage}{\columnwidth}
\begin{center}
\begin{minipage}{0.22\columnwidth}
\begin{flushleft}(a).\end{flushleft} 
\begin{center}\vspace{-0.3cm}
\includegraphics[width=\textwidth]{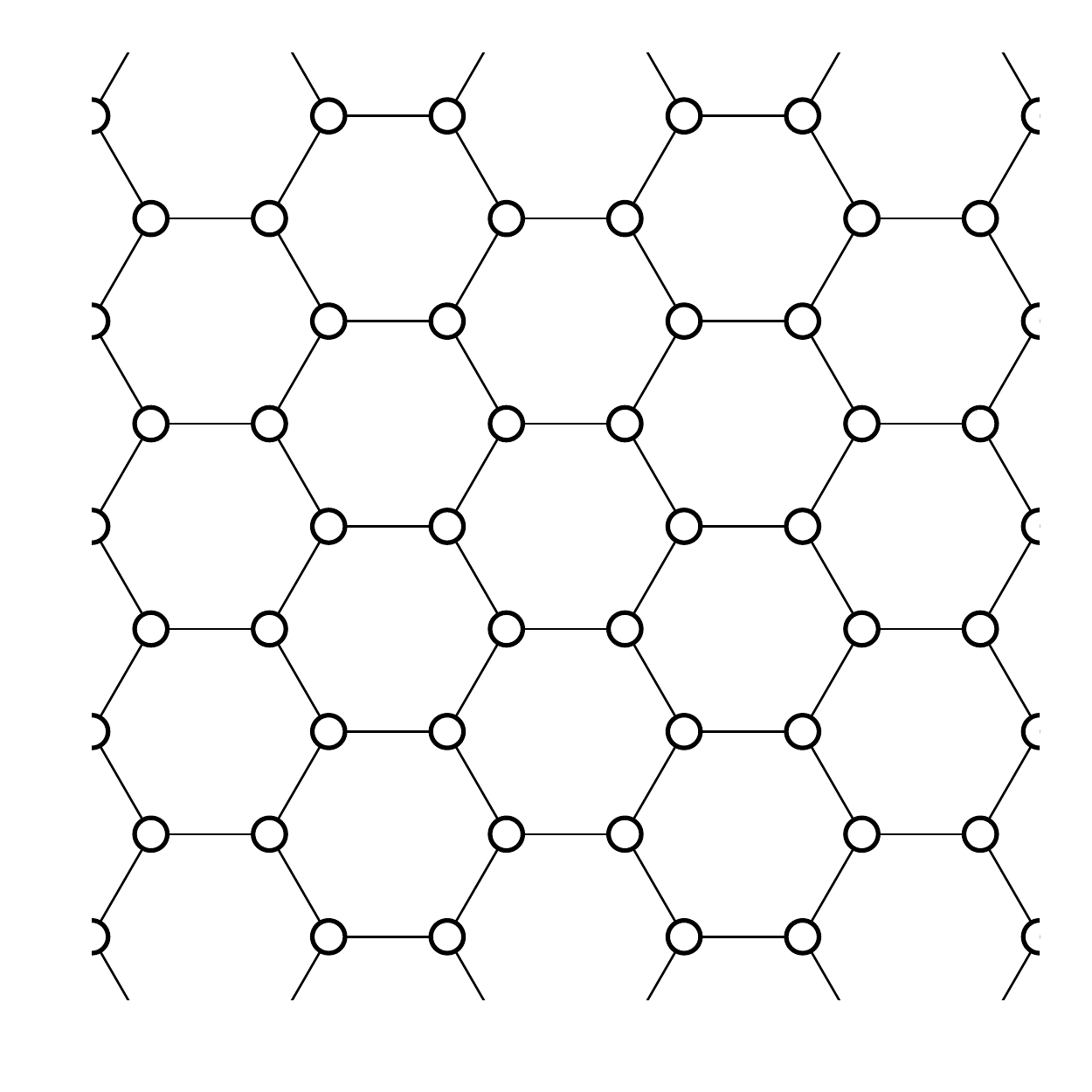}
$d=3$
\end{center}
\end{minipage}
\begin{minipage}{0.22\columnwidth}
\begin{flushleft}(b).\end{flushleft} 
\begin{center}\vspace{-0.3cm}
\includegraphics[width=\textwidth]{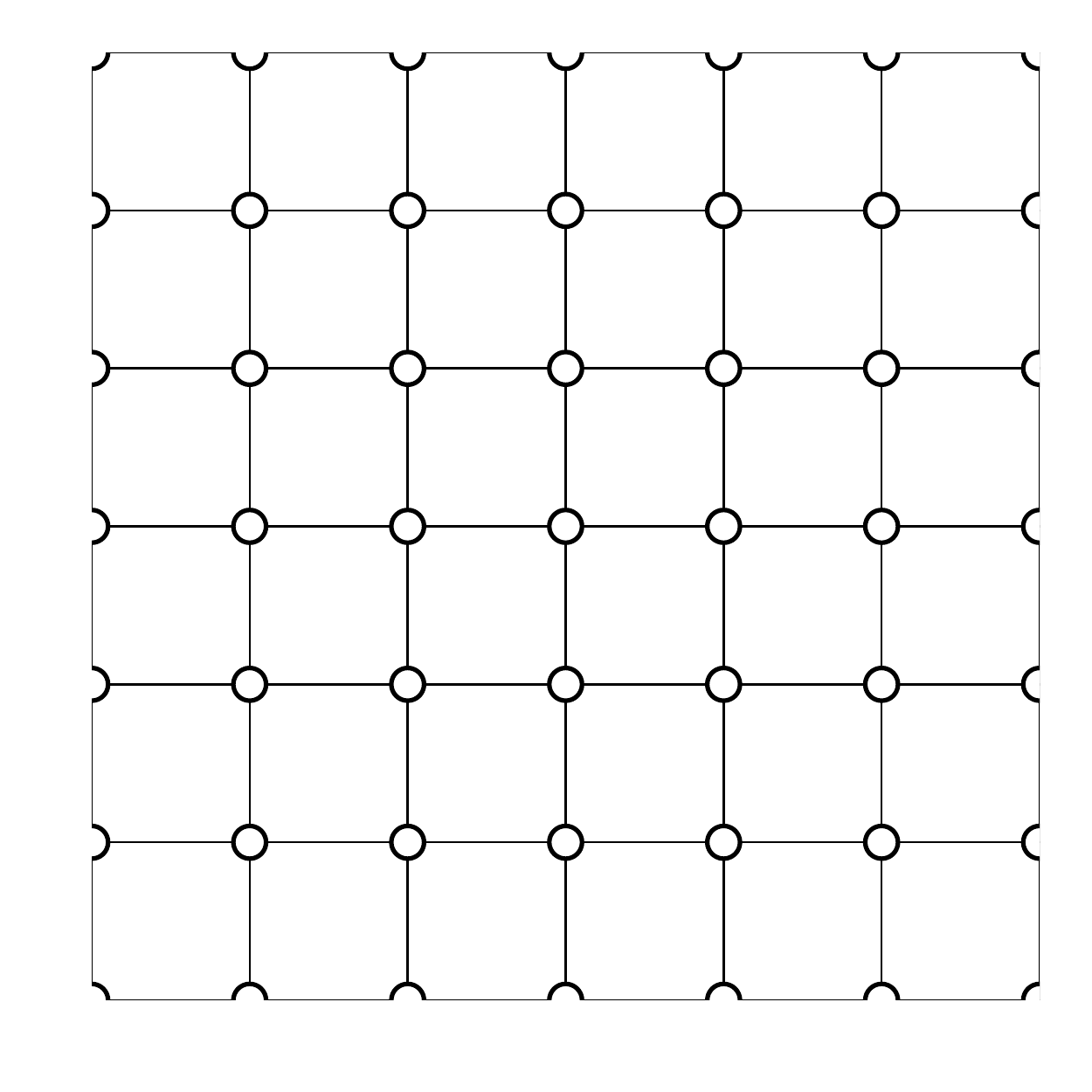}
$d=4$
\end{center}
\end{minipage}
\begin{minipage}{0.22\columnwidth}
\begin{flushleft}(c).\end{flushleft} 
\begin{center}\vspace{-0.3cm}
\includegraphics[width=\textwidth]{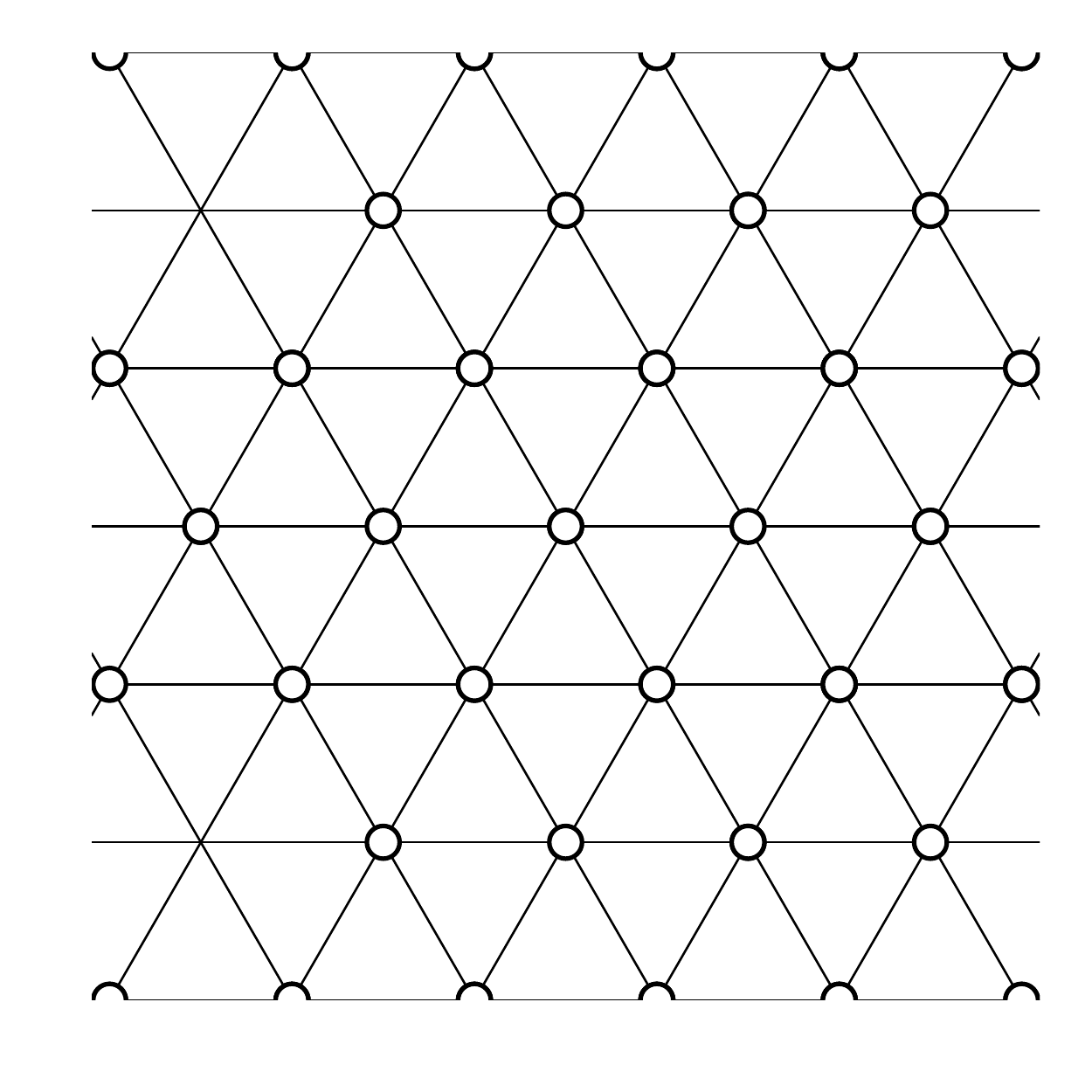}
$d=6$
\end{center}
\end{minipage}
\begin{minipage}{0.22\columnwidth}
\begin{flushleft}(d).\end{flushleft} 
\begin{center}\vspace{-0.3cm}
\includegraphics[width=\textwidth]{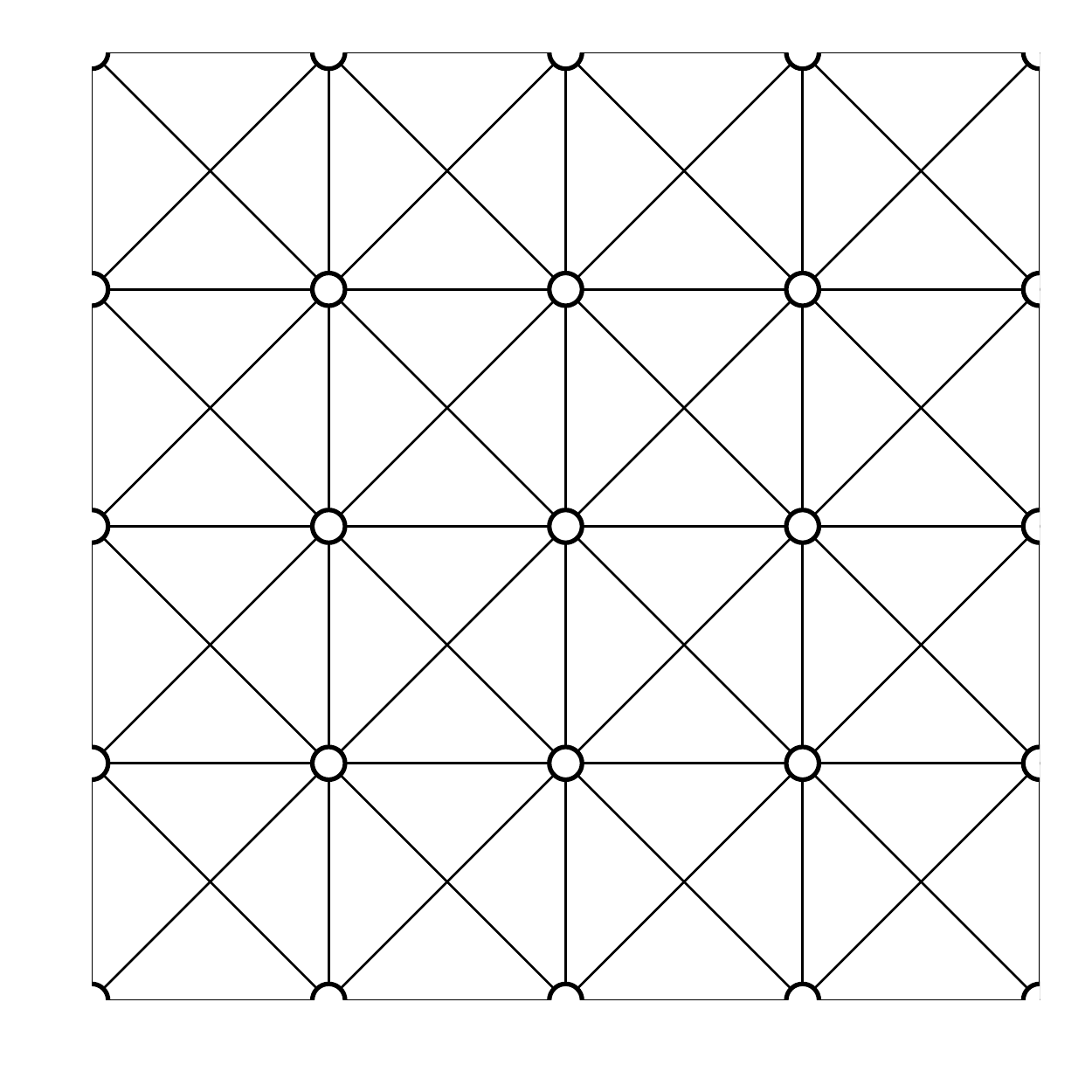}
$d=8$
\end{center}
\end{minipage}
\makeatletter\def\@captype{figure}\makeatother\caption[Four lattices are investigated in the simulations of Chase-Escape]{\label{fig:lattice} Four types of 2D lattices are investigated in the simulations of Chase-Escape}
\vspace{0.4cm}
\end{center}
\end{minipage}
\end{figure}

Let $d$ be
the degree of the lattice, i.e., the number of nearest neighbors for
each vertex. On the four types of lattices we are interested in, the
degrees are $d=3, 4, 6, $ and $8$ for hexagonal, square, triangular,
and 8-directional lattices, respectively (Fig. \ref{fig:lattice}a-d).

Similar to the Richardson, Escape and Competition models
\cite{Richardson:1973br, Kordzakhia:2005ik, Kordzakhia:2005vu}, the
Chase-Escape model can be constructed from a percolation structure\cite{MR0488377}. 
Consider the special case $\lambda_{R} =
\lambda_{B} = \lambda$ first (i.e., $p=1$). To build a percolation
structure on a lattice of degree $d$, we associate independent,
rate $\lambda d^{-1}$ Poisson point processes to all the directed edges $x
\rightarrow y$, where $x$ and $y$ are vertices. Next, a timeline is
drawn on top of each vertex with the marks at the occurrence times
$T_{i}^{x\rightarrow y}$ of the Poisson process for edge $x\rightarrow
y$. At each mark of the occurrence time $T_{i}^{x \rightarrow y}$, we
draw an arrow from  $x$ to $y$.  A \textit{directed path}
on the percolation structure can be formed by traveling along the
timeline of any vertex $x$ and making a jump to vertex $y$ at the
occurrence time $T_{i}^{x \rightarrow y}$, and a
\textit{voter-admissible} path as a directed path that does not
encounter any inward arrows.  For the Richardson model, let
$Z(0)$ be the set of red particles at time $t=0$. A version $Z(t)$ of the red
particles process can be constructed by setting
$Z(t)$ to be the set of all vertices $y$ such that there exists a
directed path on the percolation structure, starting from some $x \in
Z(0)$ and ending at $y$. For the Chase-Escape model, define the
configuration at time $t$ to be $Z(t) = R(t) \cup B(t)$, where $R(t)$
and $B(t)$ are the set of the red and blue particles at time $t$,
respectively. We first remove all the arrows from a red site to a blue
site, or from a blue site to an empty site. Then $B(t)$ is the set of
all vertices $y$ such that there exists a directed path from $x \in
B(0)$ to $y$, and $R(t)$ is the set of all vertices $y$ such that
there is a voter-admissible path starting from any $x \in R(0)$ and
terminating at $y$. For more general case in which $\lambda_{R}$ and
$\lambda_{B}$ may not be equal, we randomly remove the arrows
initiated at the particles of the slower type with probability $
\min(\lambda_{R}, \lambda_{B})/\max(\lambda_{R}, \lambda_{B})$.
\subsection{\label{subsec:numsimu} Simulation Algorithm} In our
simulations, we vary $p=\lambda_{R}/\lambda_{B}$ in the range $[0,
1]$, so that the red particles spread no faster than the blue ones, because
smaller simulations show that $\lambda_{R} = \lambda_{B}$ is already
fast enough for the red particles to escape from extinction with a
positive probability. As the main purpose of this study is to find 
the critical region for mutual survival, the exact occurrence time for each
colonization event is not important. Therefore, we will simulate the
Chase-Escape model for each $p \in [0, 1]$ using discrete time
steps. The algorithm of the simulations is as follows: at each time step
\begin{enumerate}[(1).]
\item pick a particle $x$ at random; 
\item if $x$ is red, draw a random number $X$ from a uniform-$[0, 1]$ distribution; 
\begin{enumerate}[({2}a).]
\item if $X >p$, proceed to the next time step;
\item if $X \le p$, pick at random one of the $d$ adjacent sites $y$. If $y$ is empty, colonize that site with a duplicate of the red particle; otherwise proceed to the next time step;
\end{enumerate}
\item if $x$ is blue, pick at random one of the $d$ adjacent sites $y$. If $y$ is red, change its color to blue; otherwise proceed to the next time step.
\end{enumerate}

It is not feasible to run the simulation forever on the entire infinite lattice.
 However, we can approximate the event of mutual survival by 
the event that red particles survive until hitting the boundary of a large box. 
The final configurations on these
lattices should be closer and closer to the expected outcomes on an
infinite lattice (with the same parameter settings). In our
simulations, we vary $p$ from 0.1 to 1.0 for six levels of box
sizes, ranging from $250 \times 250 $ to $8000 \times 8000$. A simulation
is terminated whenever particles reach the boundary of the lattice or the red 
particles are wiped out.
For each value of $p$, the probability of mutual survival is estimated from 500
realizations for every combination of the lattice type and box size.

We observe a sudden increase in the probability of mutual survival
from almost zero near some value $p_{c}$
(Fig. \ref{fig:psurvival:square}, more details in
Sec. \ref{subsec:coexist}), which indicates a phase transition.  The
value of $p_{c}$ is estimated and reported for each type of lattice
(Table \ref{table:pSurvivalall}).  We further investigate the growth
behavior and the geometric features of the final particle
configurations, which gives additional evidence for the criticality of 
$p_{c}$ and the phase transition.

\section{\label{sec:result}Results}
\subsection{\label{subsec:coexist} Mutual Survival and Phase
Transition} 
\begin{figure}[ht]
\begin{center}
\begin{minipage}{\columnwidth}
\includegraphics[width=\textwidth]{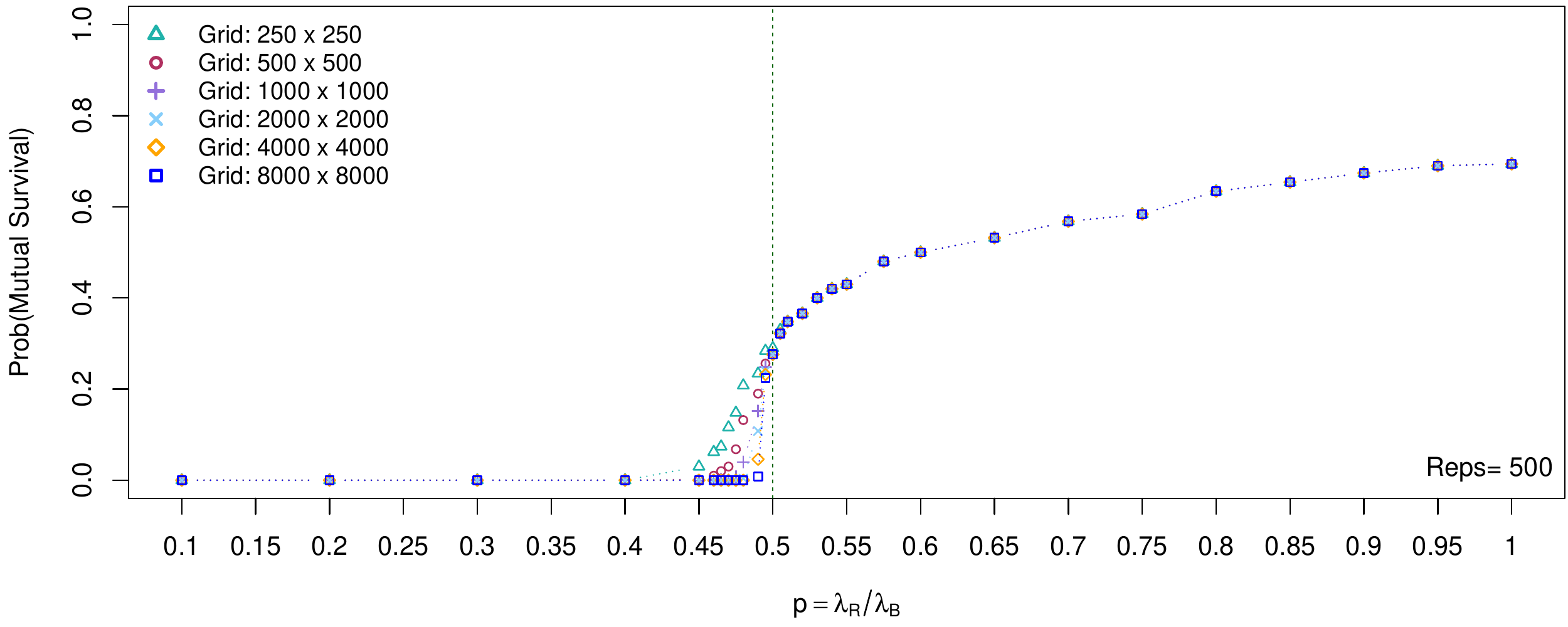}
\caption[The probability of mutual survival on square lattices for each grid size and relative growth rate. ]{\label{fig:psurvival:square} (color online) The probability of mutual survival on a square lattice for each grid size and relative growth rate. Initial configuration: $R(0) = \{(-1, 0)\}$ and $B(0) = \{(0, 0)\}$ }
\end{minipage}
\end{center}
 \end{figure}

\begin{figure}[ht]
\begin{center}
\begin{minipage}{\columnwidth}
\includegraphics[width=\textwidth]{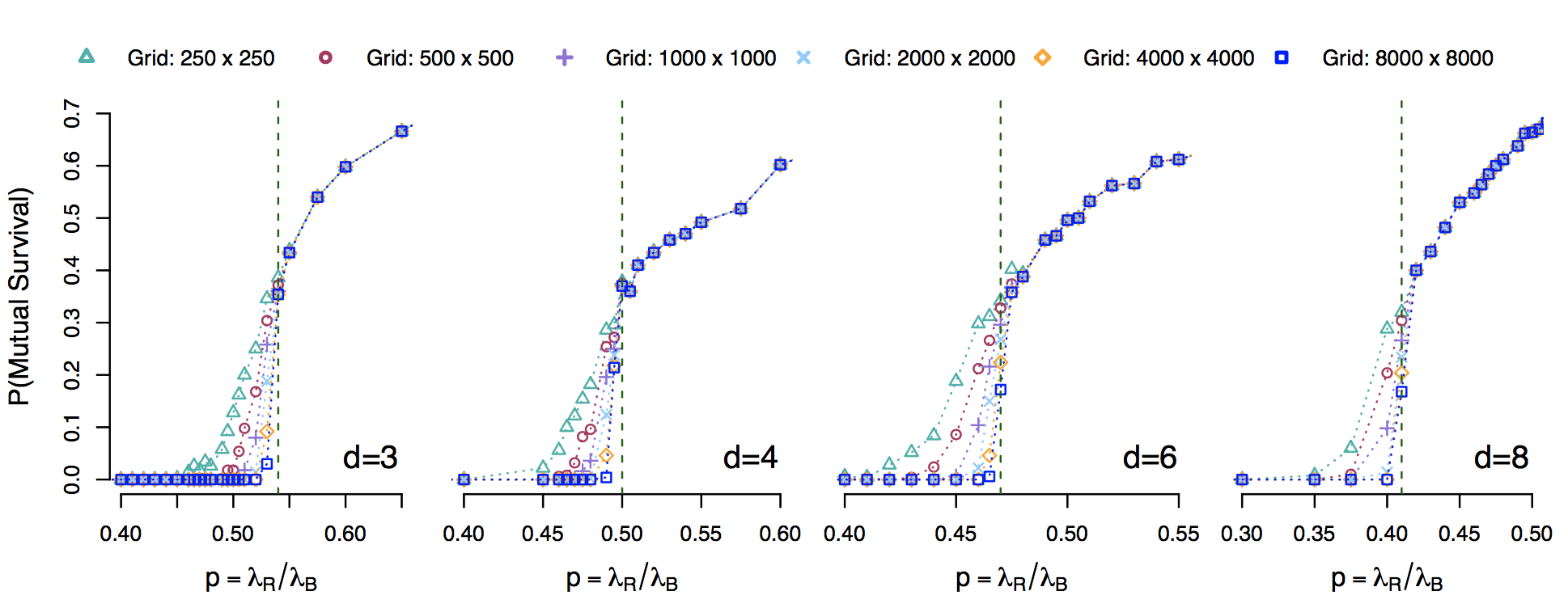}
\caption[The probability of mutual survival on four types of lattices for each grid size and relative growth rate $p$ around the phase transition. ]{\label{fig:psurvival:all} (color online) The probability of mutual survival on four types of lattices for each grid size and relative growth rate $p$ around the phase transition. Initial configuration of the square lattice: $R(0) = \{(-1, 0), (1, 0), (0, 1), (0, -1)\}$ and $B(0) = \{(0, 0), (-1, 1), (-1, -1), (1,1), (1, -1)\}$ }
\end{minipage}
\end{center}
\end{figure}

On the square lattice, when the relative growth rate $p$
is very small, the chance of mutual survival is close to zero.  As $p$
increases from $0$ to $1$, the probability of mutual survival does not
change much until at some value $p_{c}$ it suddenly increases from near
zero to a positive value. After that the probability of mutual
survival continues to increase with $p$ but with a much smaller
slope. (Fig. \ref{fig:psurvival:square}). The change of the mutual survival probability is sharper on larger lattices: for the largest grid size, $8000 \times 8000$ (blue squares in Fig. \ref{fig:psurvival:square}), the transition is the sharpest, from $\mathbb P(\text{mutual survival})=0.006$ when $p=0.49$ to $\mathbb P(\text{mutual survival})=0.412$ when $p=0.51$.  Fig. \ref{fig:psurvival:square} strongly suggests that the phase transition occurs at $p_{c} \approx 0.5$, independent of the initial configurations of red and blue particles (see Fig. \ref{fig:psurvival:square} and Fig. \ref{fig:psurvival:all}, second panel from the left, where the two initial configurations are different).

The same pattern of the mutual survival probability is observed (Fig. \ref{fig:psurvival:all}) for the other three types of lattices as well, but the values $p_{c}$ at which the jumps take place are different (Table \ref {table:pSurvivalall}).  These critical values are monotonically decreasing with the degree of the lattices (Table \ref{table:pSurvivalall}).


\begin{table}[ht] 
 \makeatletter\def\@captype{table}\makeatother\caption{\label{table:pSurvivalall} Critical values for the relative growth rates on four different types of lattices}\br 
\begin{ruledtabular}
     \begin{tabular}{cccccc}
Lattice Type && Hexagon & Square & Triangle & 8-direction\\ 
 Degree, $d$ &&  $d=3$ & $d=4$ & $d=6$ & $d=8$\\ \hline
 $p_{c}$ (approx.) & & 0.54 & 0.50 &0.47 & 0.41 \\
     \end{tabular}
     \end{ruledtabular}
\end{table}

\begin{figure}[h]
\begin{center}
\includegraphics[width=0.9\columnwidth]{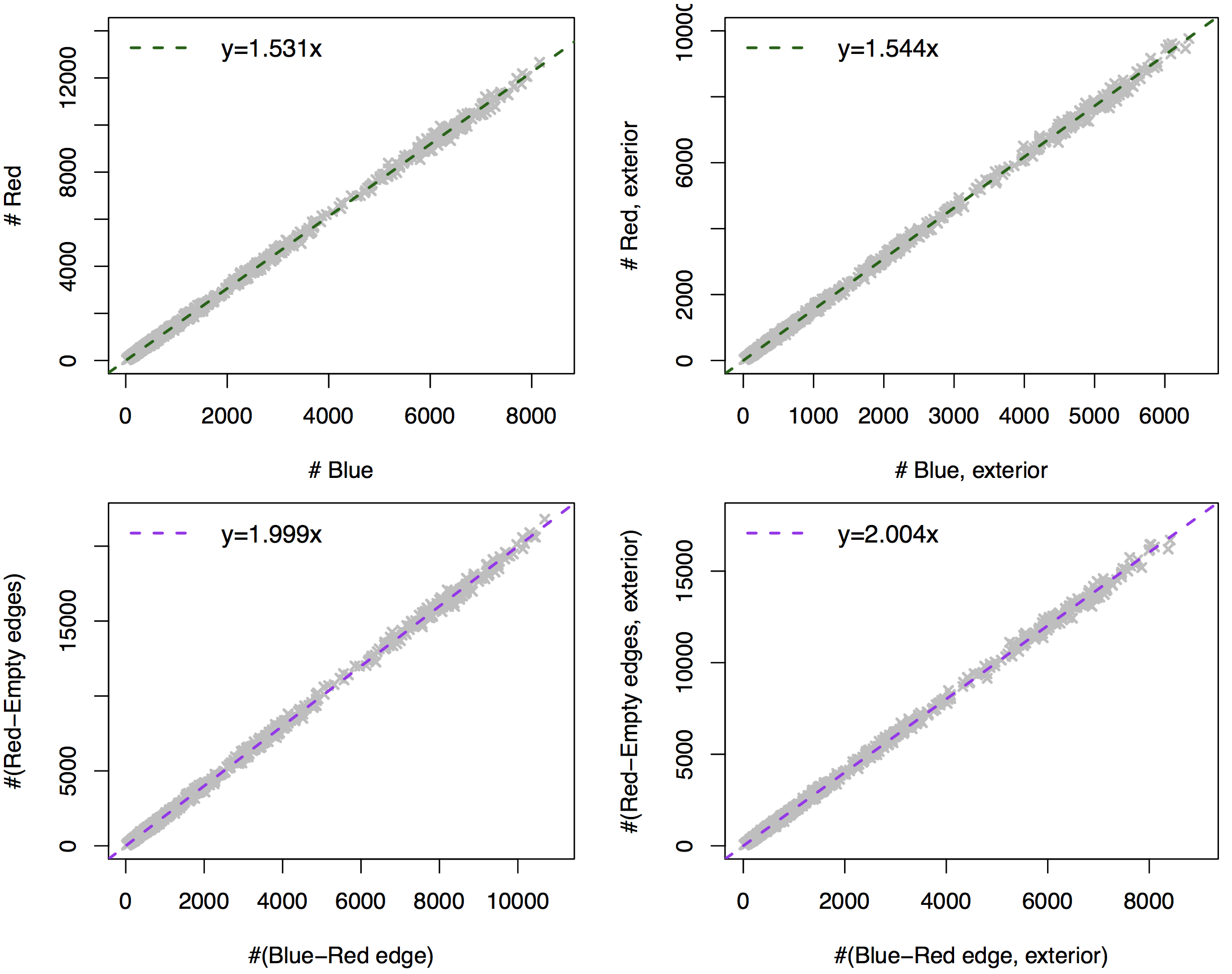}
\makeatletter\def\@captype{figure}\makeatother\caption{\label{fig:pc:point-edge} (color online) The number of active particles (upper two panels, left: all active particles; right: active exterior particles only) and the number of active edges (lower two panels, left: all active edges; right: active exterior edges only) on square lattices. The dotted line in each plot is a linear regression of the points. }
\end{center}
\end{figure}

Because red particles can only colonize empty sites, the total number
of red-empty edges is a measure of the space where red particles can
grow. To compare the spaces where red and blue particles can colonize
in the next time step, we count the number of red-empty edges and the
number of blue-red edges (Fig. \ref{fig:pc:point-edge}).  At the value
$p=0.5$ (near criticality), our simulations show that the red-empty
edges are about twice as numerous as the blue-red edges. In addition,
the number of active red particles is about 1.5 times the number of
active blue particles. (By ``active'' we mean the particles that are
able to grow in the next time step, i.e., an active red (resp. blue)
particle must be adjacent to at least one empty (resp. red) site.)
 
For $p=.5$, we plot the number of red-empty edges against the blue-red edges and
active red particles against the active blue particles in Figure
4. These plots show a clear linear dependence.  For values of $p$ not
near $.5$, such linear relationships disappear:  the ratios, $\frac{\text{number of red-empty edges}}{\text{number of blue-red edges}}$ and $\frac{\text{number of active red particles}}{\text{number of active blue particles}}$ vary considerably across  simulations (data not shown). 

 \subsection{\label{subsec:fractal}  Fractal nature of the growing disk}      
To exhibit the apparent fractal character of the colonized regions
near the critical value, we report here on the values of
several suggestive numerical  quantities of the final configurations.
We limit our discussion here to  the square lattice case, which
illustrates the phenomena found on all lattices.

\begin{figure}[h]
\begin{center}
\begin{minipage}{\columnwidth}
\begin{minipage}{0.32\textwidth}
\begin{flushleft}(a). $p= 0.50$\end{flushleft} \vspace{-0.4cm}
\includegraphics[width=\textwidth]{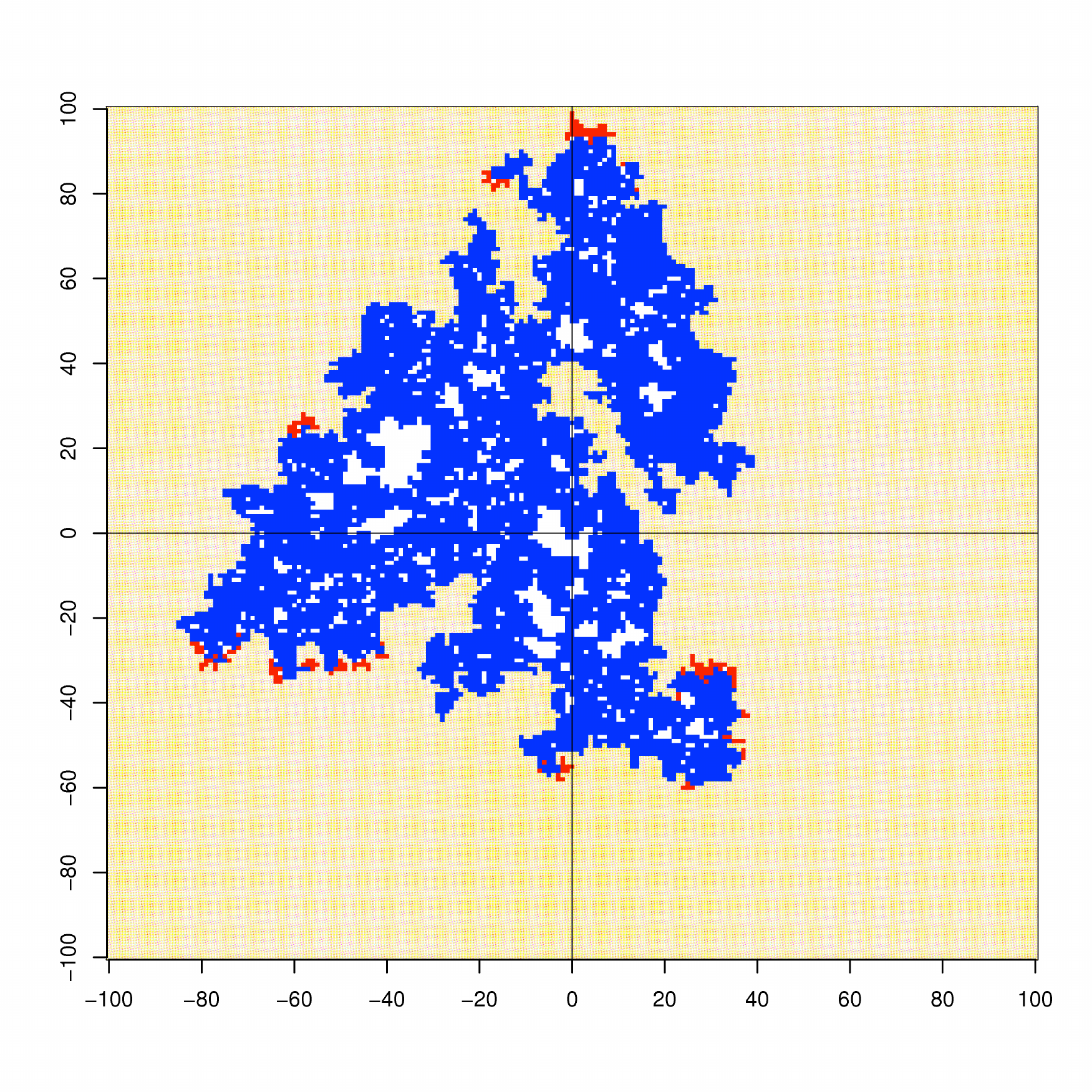}
\end{minipage}
\begin{minipage}{0.32\textwidth}
\begin{flushleft}(b). $p= 0.75$\end{flushleft}  \vspace{-0.4cm}
\includegraphics[width=\textwidth]{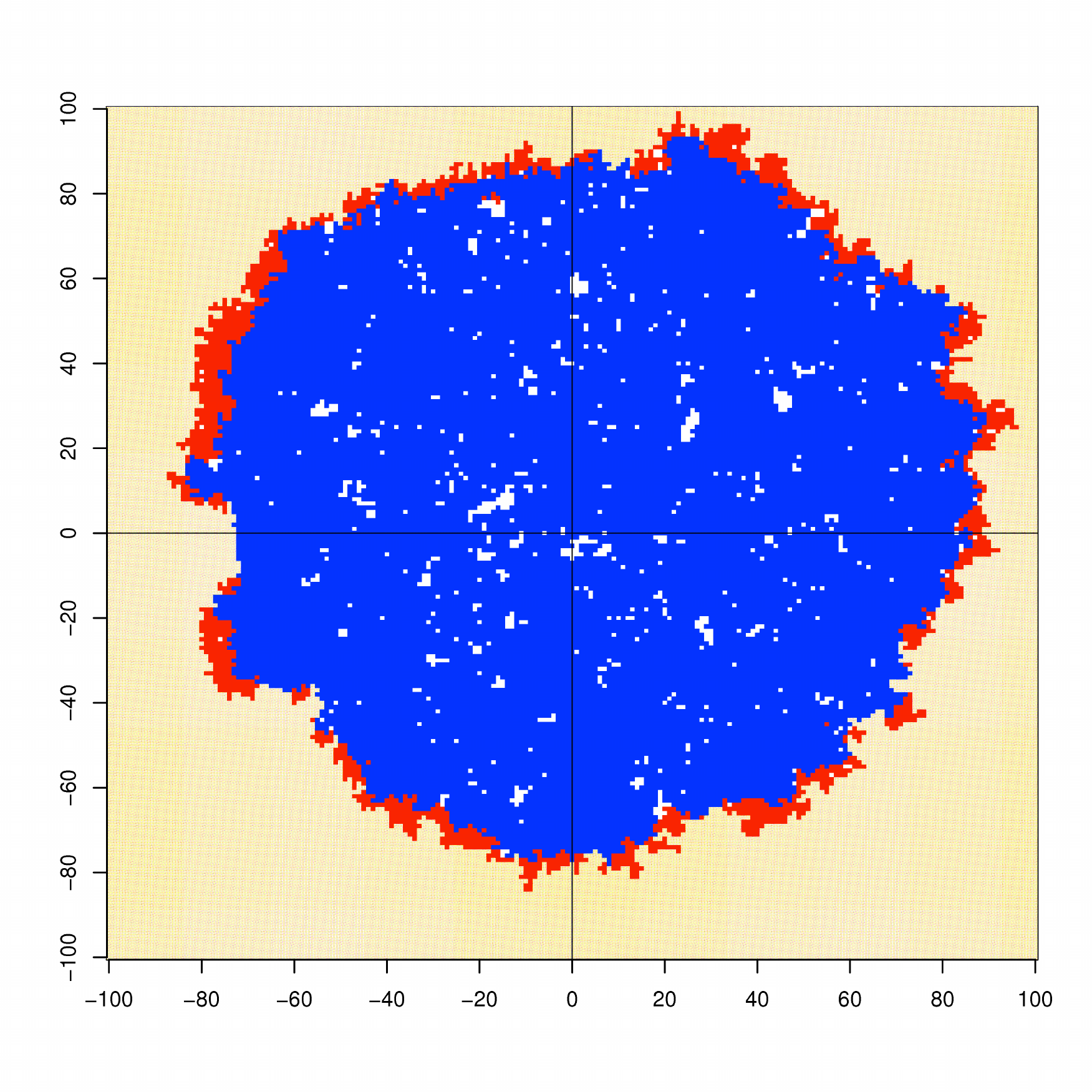}
\end{minipage}
\begin{minipage}{0.32\textwidth}
\begin{flushleft}(c). $p=1.00$\end{flushleft}  \vspace{-0.4cm}
\includegraphics[width=\textwidth]{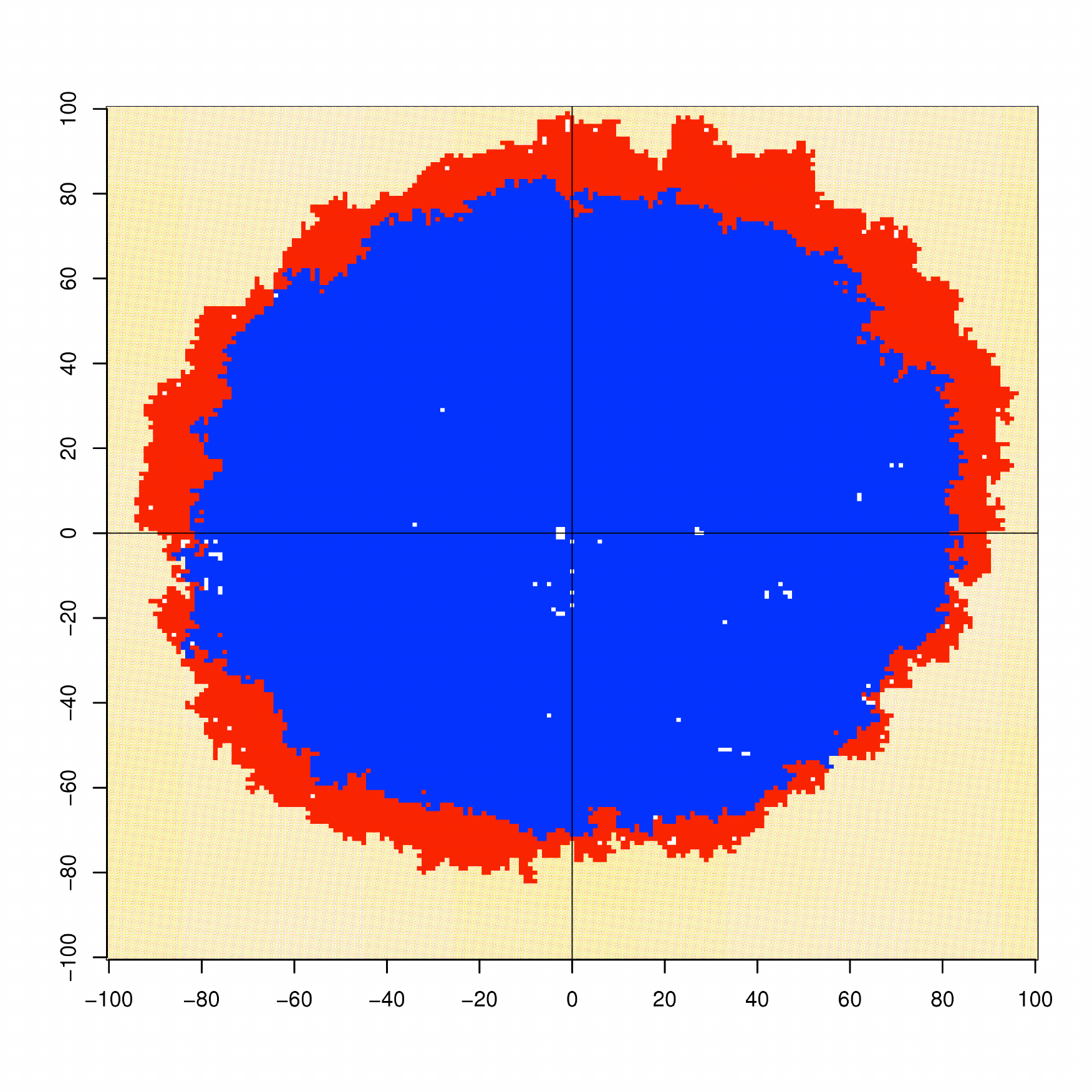}
\end{minipage}
\caption[The shapes of the Chase-Escape model at different relative growth rates]{\label{fig:2DCEShapes} (color online) The shapes of the Chase-Escape model at different relative growth rates. The initial configuration is the same as that in Fig. \ref{fig:psurvival:all}}
\end{minipage}
\end{center}
\end{figure}

In Fig. \ref{fig:2DCEShapes}, we plot the final configurations of red
and blue particles at $p=0.5, 0.75$ and $1.00$ on a $200 \times 200$
lattice. The plots show that when the relative growth rate of red to
blue is closer to $0.5$ (our conjectured threshold), the geometry of the
final configuration becomes more fractal. In contrast, the geometry is more regular,
and resembles  the Richardson shape \cite{Durrett:1981vs} when
$p>.5$. Keep in mind  that red
particles can only survive on the outer boundary, so the inner area
consists of blue particles and empty sites. When $\lambda_{R}$ is
sufficiently larger than $0.5 \lambda_{B}$, red particles can form a
thin layer on the outer boundary of the blue disk, whereas if
$\lambda_{R} / \lambda_{B} = 0.5 $, red particles can only survive at
the tips of the blue branches.

\begin{figure}[ht]
\vspace{0.2cm}
\begin{center}
\begin{minipage}{\columnwidth}
\begin{minipage}{0.24\textwidth}
\begin{flushleft}(a).\end{flushleft} \vspace{-3mm}
\includegraphics[width=\textwidth]{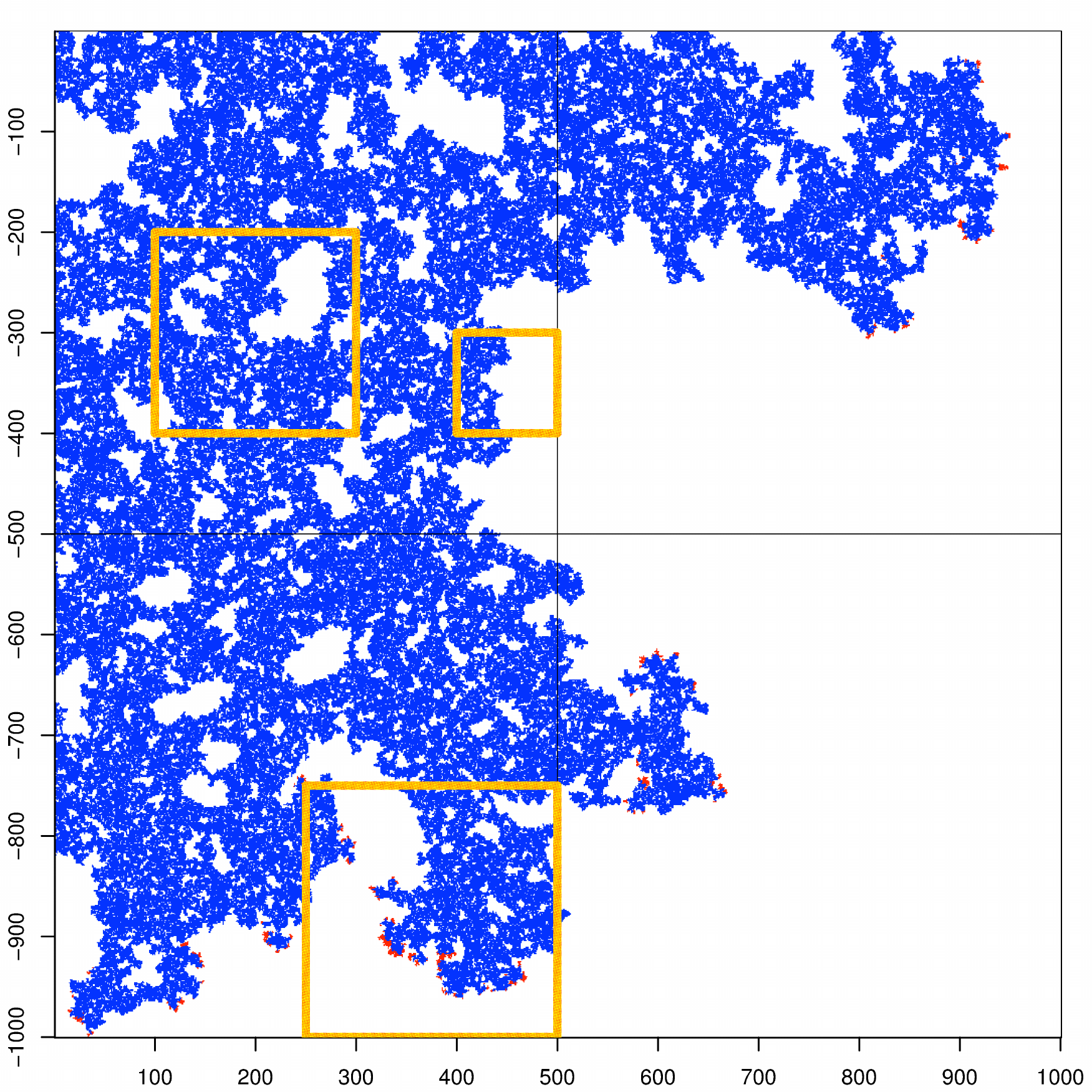}
\end{minipage}
\begin{minipage}{0.24\textwidth}
\begin{flushleft}(b).\end{flushleft}  \vspace{-3mm}
\includegraphics[width=\textwidth]{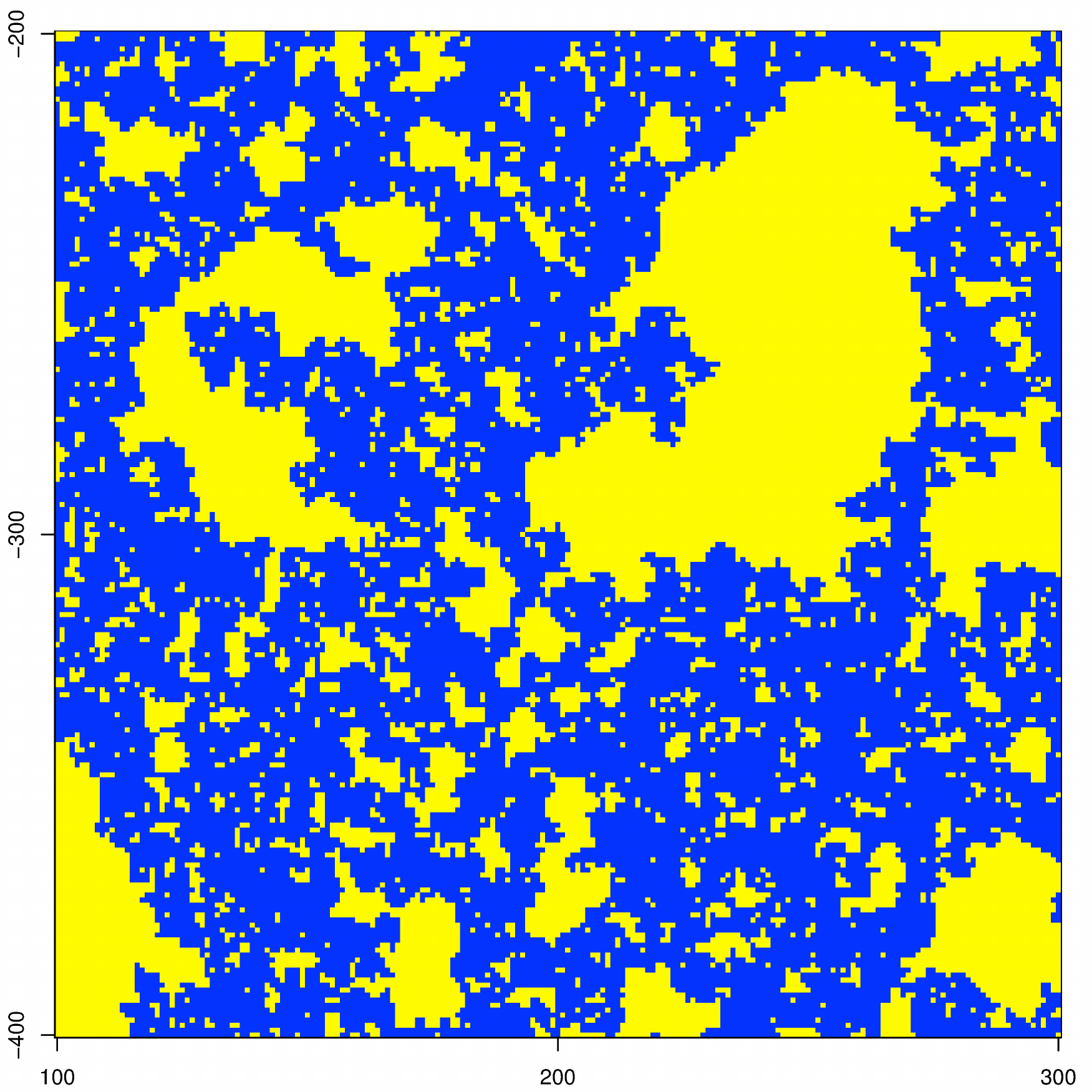}
\end{minipage}
\begin{minipage}{0.24\textwidth}
\begin{flushleft}(c).\end{flushleft} \vspace{-3mm} 
\includegraphics[width=\textwidth]{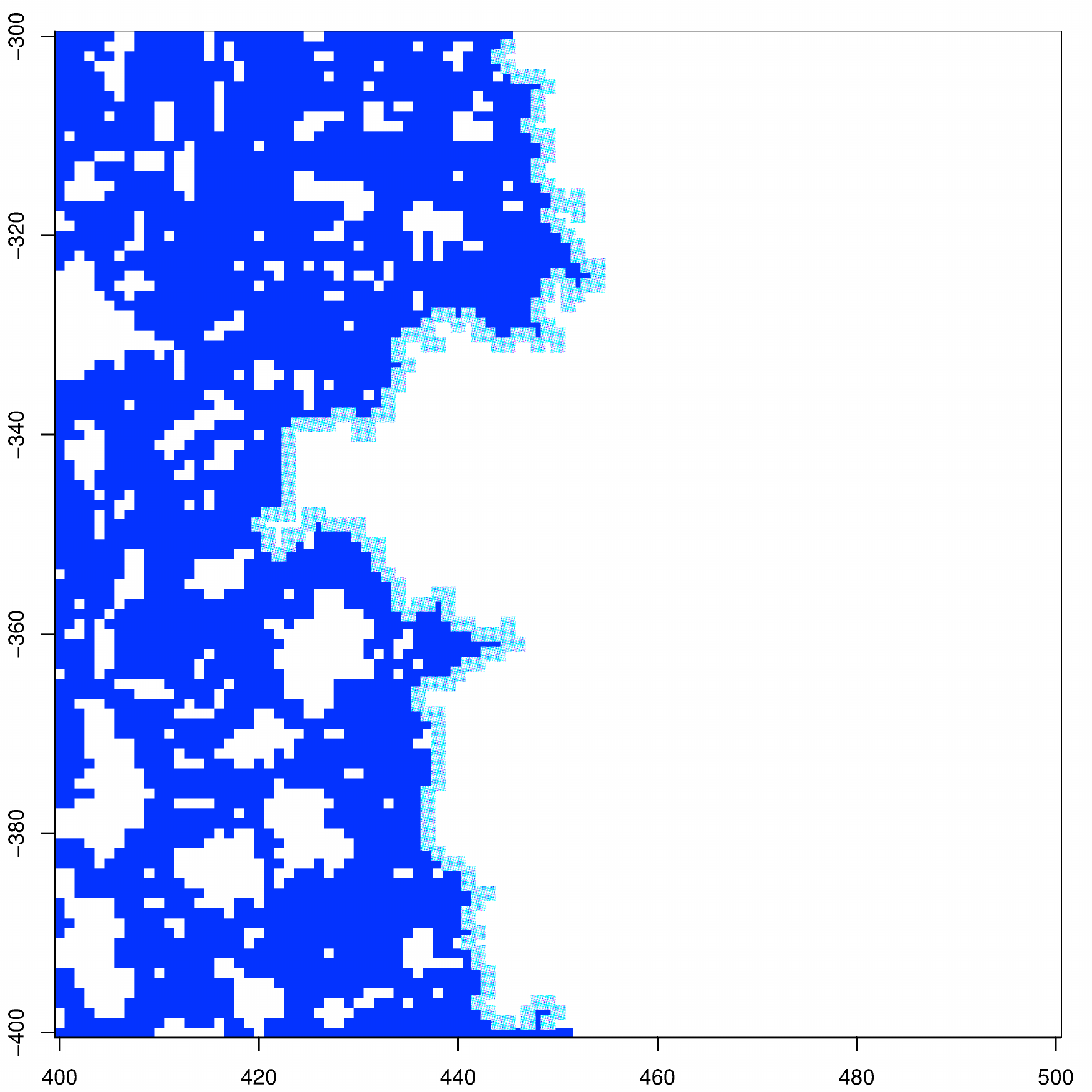}
\end{minipage}
\begin{minipage}{0.24\textwidth}
\begin{flushleft}(d).\end{flushleft} \vspace{-3mm}
\includegraphics[width=\textwidth]{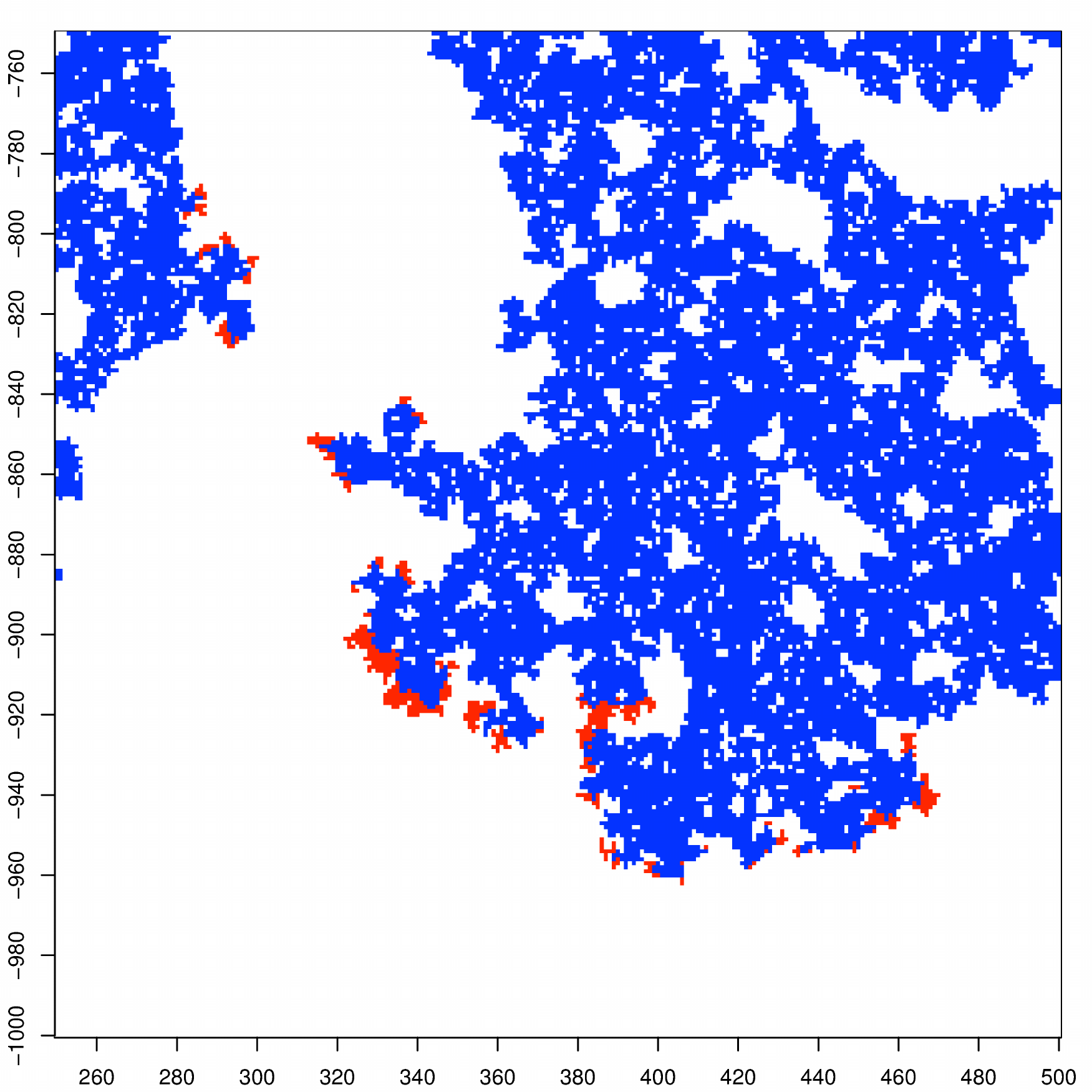}
\end{minipage}
\makeatletter\def\@captype{figure}\makeatother\caption[Critical growing behavior ($p=0.5$) on a $2000 \times 2000$ square lattice]{\label{fig:shape}  \label{fig:2000lowerdetail}(color online) Critical growing behavior ($p=0.5$) on the square lattice of size $2000 \times 2000$. (a). configurations of red and blue particles in the low-right quadrant, i.e., $[0, 1000] \times [-1000, 0]$; (b). a large number of unfilled area (highlighted in yellow); (c).curly boundary (marked in light blue); (d). local survival of the red particles. }
\end{minipage}
\end{center}
\end{figure}

We do not show plots of the final configurations when $p$ is far below
$p_{c}$ because in those simulations the red population dies out
quickly and the area filled by the particles during the simulation is
too small to see any limiting patterns or structures. (In fact, the
simulation results reported in Fig. \ref{fig:psurvival:square} show
that probability of mutual survival is near zero when $p < 0.45$ for
all grid sizes.) A closer look at  three local areas of the final
configuration on a $2000\times 2000$ square lattice, shown in
Fig. \ref{fig:2000lowerdetail}a,  confirms that (1) the final shape is
extremely irregular, characterized by a large number of holes
(unoccupied interior sites) of various sizes
(Fig. \ref{fig:2000lowerdetail}b), and highly irregular  boundary
(Fig. \ref{fig:2000lowerdetail}c); and (2) if red particles 
survive in the simulations, most red buds are attached to blue
tips (Fig. \ref{fig:2000lowerdetail}d).

\begin{figure}[ht]
\begin{center}
\begin{boxedminipage}{0.315\columnwidth}
\centering
\includegraphics[width=\textwidth]{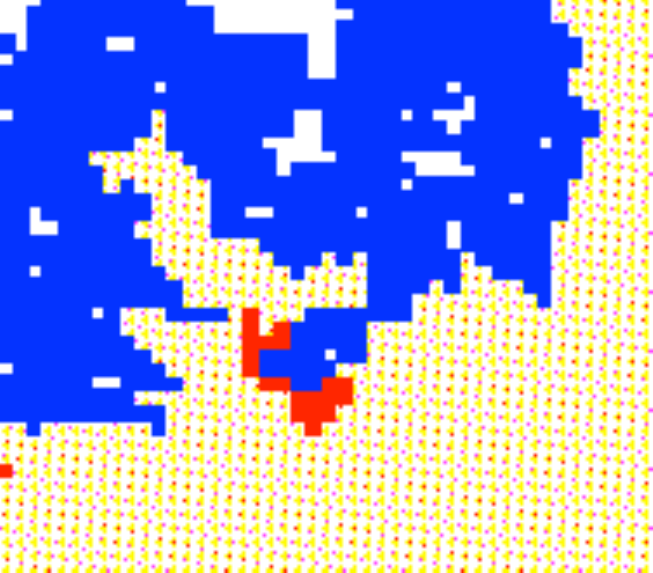}
\end{boxedminipage}
\hskip 0.1cm
\begin{boxedminipage}{0.315\columnwidth}
\centering
\includegraphics[width=\textwidth]{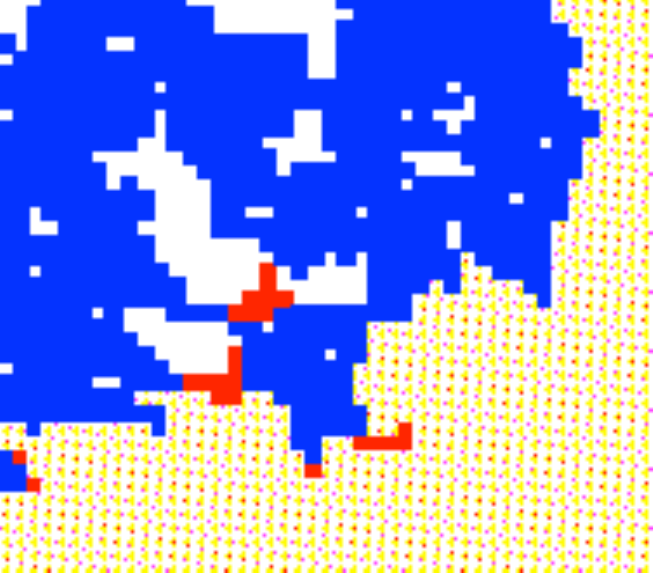}
\end{boxedminipage}
\hskip 0.1cm
\begin{boxedminipage}{0.315\columnwidth}
\centering
\includegraphics[width=\textwidth]{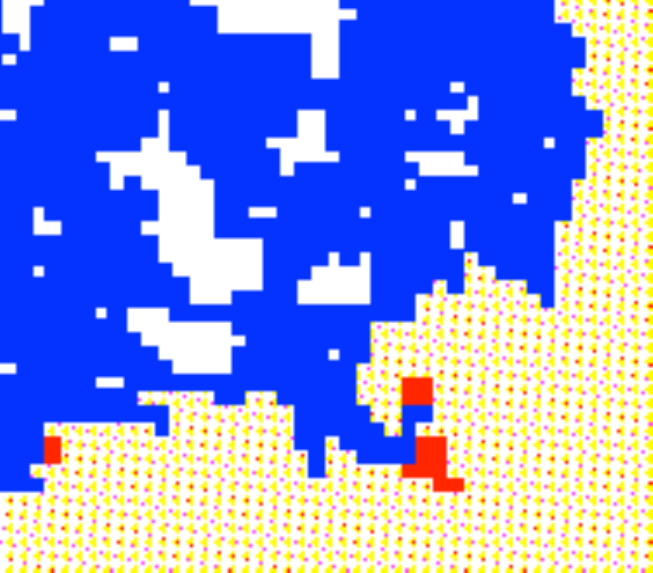}
\end{boxedminipage}
\vspace{0.2cm}
\makeatletter\def\@captype{figure}\makeatother\caption[Holes formed by occasional joints of two growing branches]{\label{fig:formhole} (color online) Holes (white area) formed by occasional joints of two growing branches. } 
\end{center}
\end{figure}

We present three snapshots from a realization on a $100\times 100$ square lattice at $p=0.5$ to demonstrate how these holes are created (Fig. \ref{fig:formhole}). Occasionally, red particles may expand locally and wind themselves to connect to a distant part of the occupied area, leaving an empty area completely surrounded by the two types of particles. If the red particles go extinct locally before this empty area gets filled up, a hole is left and it has no chance to be filled up as there are no red particles adjacent to these empty sites.

\begin{figure}[ht]
 \begin{center}
\includegraphics[width=0.47\columnwidth]{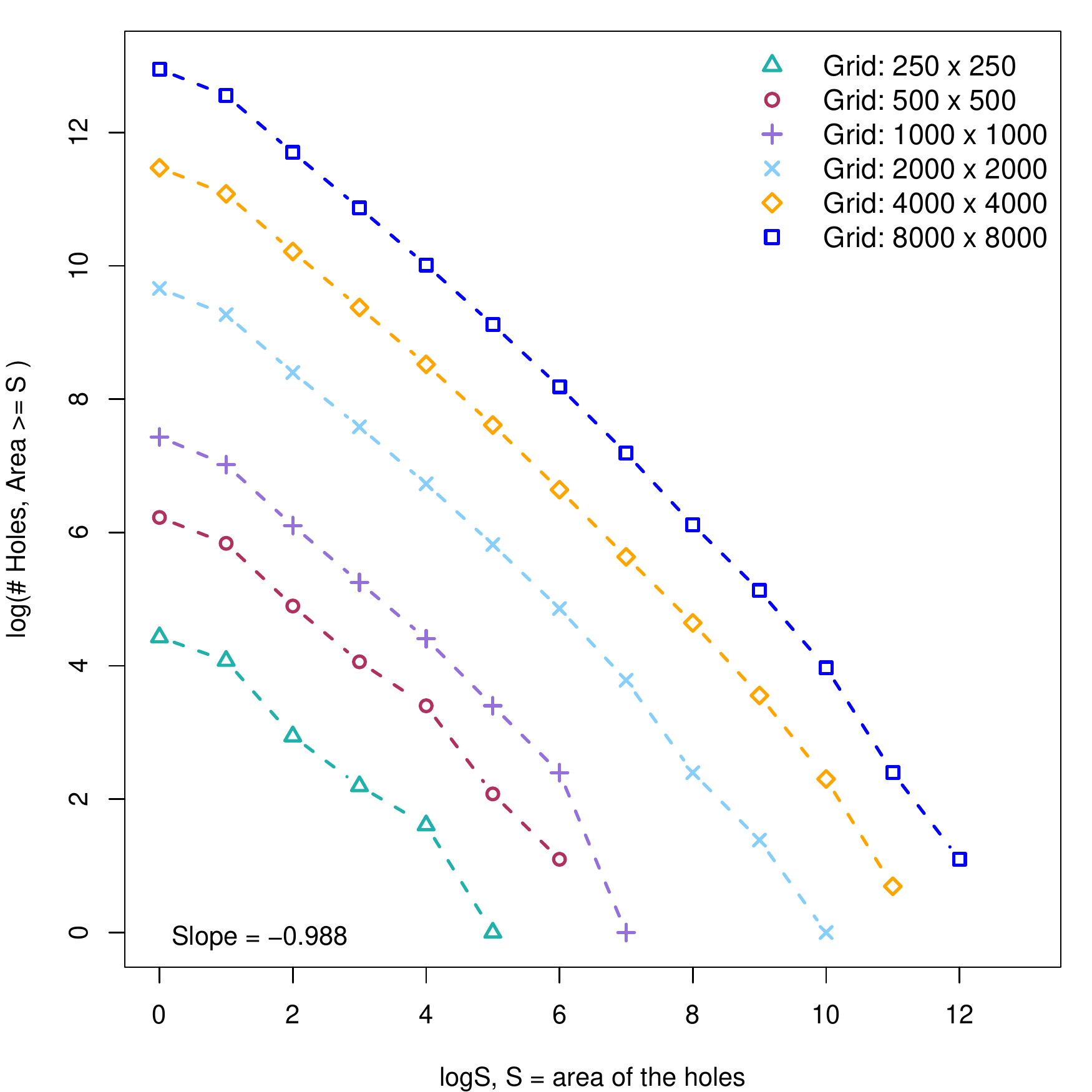}
\includegraphics[width=0.47\columnwidth]{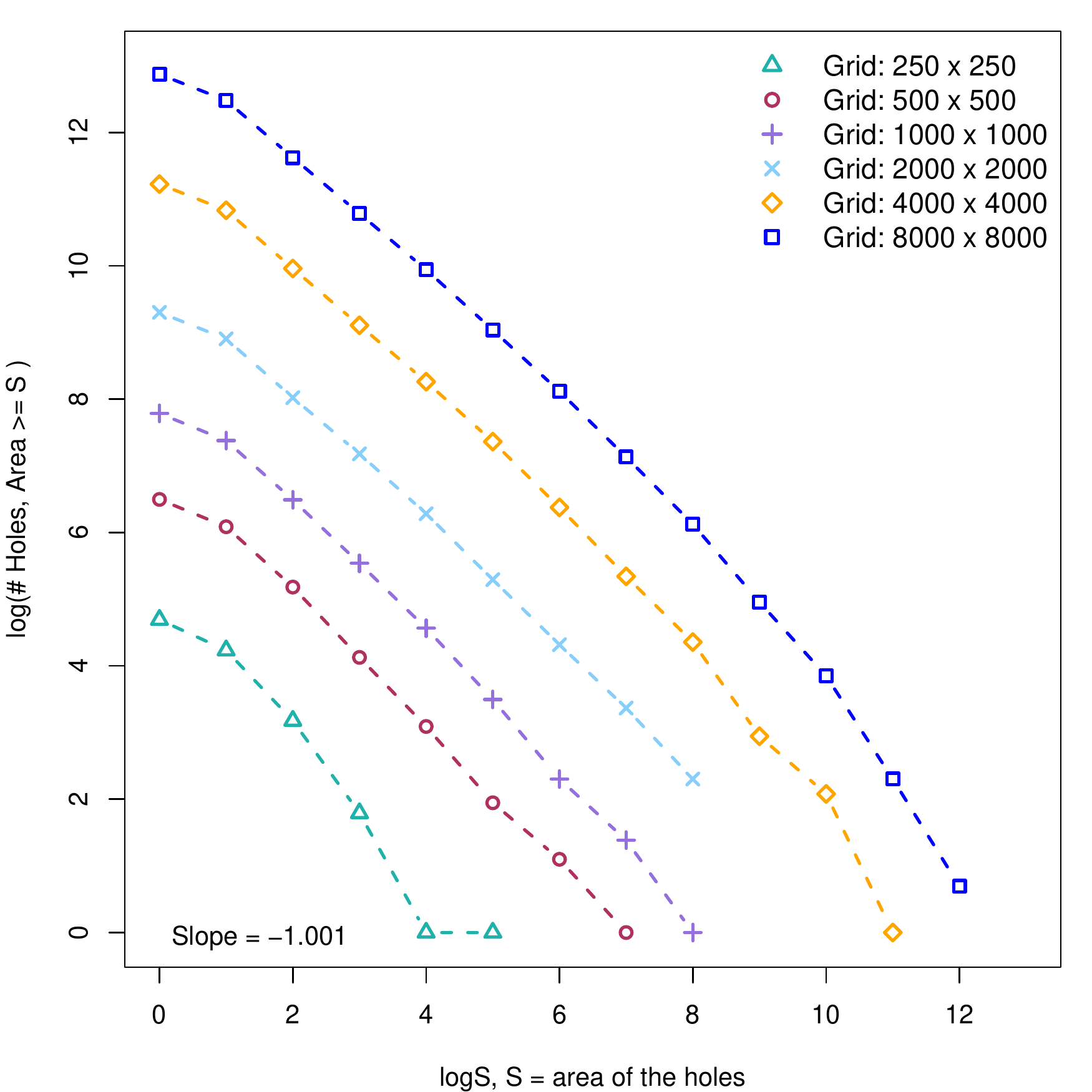}
\caption{\label{fig:hole} (color online) The areas of the holes follow a power law distribution: $N(S) \approx S^{-1}$, where $N(S)$ is the number of holes with at least an area $S$. The slope reported at the lower left corner is estimated using linear regression $\log N(S) \sim \log S$ on the largest square lattice of size $8000\times 8000$. Left and right panels are two different simulations at $p=0.5$}
\end{center}
\end{figure}

Fig. \ref{fig:2000lowerdetail}b shows that large holes are rare and
tiny holes are abundant. Let $N(S)$ denote the number of holes whose
areas are at least $S$. At $p=0.5$, 
plots of $\log N(S)$ versus $\log S$  (Fig. \ref{fig:hole}) show a
nearlhy linear relationship, which suggests the relation 
\begin{align}
N(S) \approx S^{-1}.
\end{align}
The power law relationship between $N(S)$ and $S$ is consistently
observed across  all types of lattices.

\begin{figure}[ht]
\begin{center}
\includegraphics[width=\columnwidth]{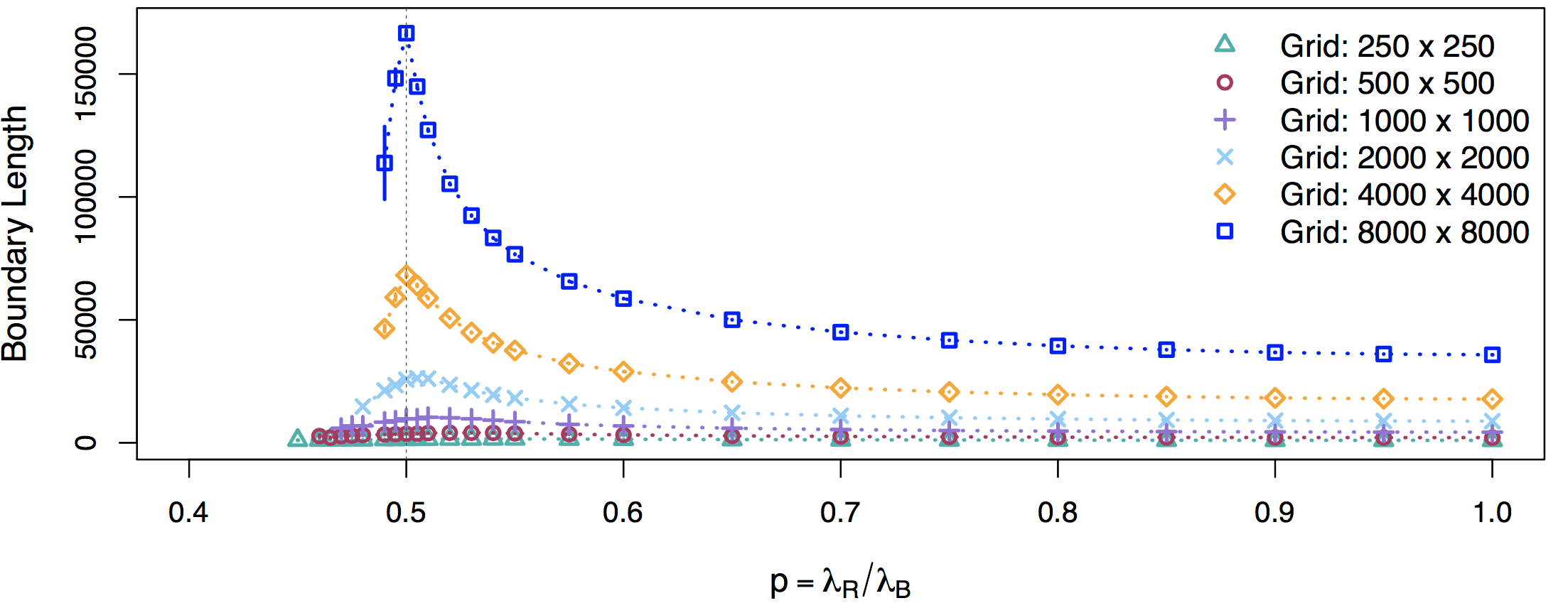}
\caption[Longest boundary length]{\label{fig:boundary} (color online) The boundary length of the final shape for all values of $p$ when red particles do not go extinct. The vertical line on each point shows the mean standard error (MSE) of the boundary length for 500 simulations.}
\end{center}
 \end{figure}

Finally, we consider the total outer boundary length, i.e., the total number of particles adjacent to empty sites that are not part of the holes (Fig. \ref{fig:boundary}). 
 There is a sharp peak of the boundary length occurring at $p=0.5$ on the square lattice, which drops very quickly as $p$ deviates from 0.5, indicating that the boundary is smoother. 

\section{\label{sec:conclusion} Concluding Remarks}

Denote by $\theta_{p}$ the probability of mutual survival when the
relative growth rate is $p$. Our simulations confirm that (1)
$\theta_{p}$ is increasing with $p$ on four types of lattices and (2)
there is a critical value $p_{c}$, whose value depends on the type of
lattice, such that if $p>p_{c}$, then $\theta_{p}>0$ and if
$p < p_{c}$, then $\theta_{p} =0$. The
critical value $p_{c}$ apparently decreases with the degree of the
lattice. This is as might have been expected, because when the degree
of the lattice is larger, red particles have more directions to escape
from being eaten by blue particles. Since the colonization times of
blue particles are independent of the times when red particles can
spread to the empty sites, when the degree is larger it is less likely
that a blue particle will expand in exactly the same direction in the
next time step after red occupies a new site, making it easier for
red particles to survive. Our simulations support the conjecture 
that red particles can survive when its colonization rate is strictly 
slower than that of the blue colonization rate on $\mathbb Z^{d}$ 
with $d \ge 2$. In particular, $p_{c}\approx 1/2$ on $\mathbb Z^{2}$. 
We find that when $\lambda_{R}/\lambda_{B} = p_{c}$, the number of red-empty edges is close to
$1/p_{c}$ times that of the blue-red edges during the entire process. This means, at $p_{c}$, empty sites are colonized at about the same rate as red sites being taken over, implying the criticality of $p_{c}$. 

At the proposed critical value, a phase transition is observed, and the shape
formed by the red and blue particles is fractal. Such phenomena are
common in related  models of statistical physics. In all the
simulations where red particles survive, we have found the most
fractal final geometry at the proposed critical value, characterized
by a great number of holes with various sizes, curliest and longest
boundary. Red particles can only survive at the tips of the growing
branches, as the interior red particles will be eventually replaced by
the blue ones.  Formation of large holes happens rarely, whereas small holes
are formed more frequently. The areas of the holes are characterized
by a power law, with an exponent approximately $-1$.



%

\end{document}